\documentclass{ws-ijmpd}

\begin{document}

\markboth{Sandro D. P. Vitenti, Mariana P. Lima, Marcelo J. Rebou\c{c}as}
{Dark Energy Equation of State and Cosmic Topology}

%
\catchline{}{}{}{}{}
%

\title{\uppercase{Dark Energy Equation of State and Cosmic Topology}}

\author{SANDRO D. P. VITENTI, MARIANA P. LIMA AND MARCELO J. REBOU\c{C}AS}

\address{Centro Brasileiro de Pesquisas F\'{\i}sicas, Rua Dr. Xavier Sigaud 150\\
22290-180 Rio de Janeiro -- RJ,
Brasil\\
vitenti@cbpf.br, penna@cbpf.br, reboucas@cbpf.br}

\maketitle

\begin{history}
\received{Day Month Year}
\revised{Day Month Year}
\comby{Managing Editor}
\end{history}

\begin{abstract}
The immediate observational consequence of a non-trivial
spatial topology of the Universe is that an observer could
potentially detect multiple images of radiating sources.
In particular, a non-trivial topology will generate pairs
of correlated circles of temperature fluctuations in the
anisotropies maps of the cosmic microwave
background (CMB), the so-called circles-in-the-sky.
In this way, a detectable non-trivial spatial topology
may be seen as an observable attribute, which can be probed
through the circles-in-the-sky for all locally homogeneous
and isotropic universes with no assumptions on the cosmological
dark energy (DE) equation of state (EOS) parameters.
We show that the knowledge of the spatial topology through
the circles-in-the-sky offers an effective way of reducing
the degeneracies in the DE EOS parameters. We concretely
illustrate the topological role by assuming a Poincar\'{e}
dodecahedral space topology and reanalyzing the constraints
on the parameters of a specific EOS
which arise from the supernovae type Ia,
baryon acoustic oscillations and the CMB plus the
statistical topological contribution.
\end{abstract}


\section{Introduction}

In the  standard cosmology, the Universe is modelled by a space-time
manifold $\mathcal{M}_4 = \mathbb{R} \times M_3$ endowed with a
locally (spatially) homogeneous and isotropic Robertson-Walker
metric
\begin{equation} \label{eq:rwmetric}
ds^2 = -c^2\,dt^2 + a^2 (t) \left [ d \chi^2 +
S_k^2(\chi) (d\theta^2 + \sin^2 \theta  d\phi^2) \right ],
\end{equation}
where $a(t)$ is the cosmological scale factor, and
$S_k(\chi)=(\chi\,$, $\sin\chi$, $\sinh\chi)$ depending on the sign
of the constant curvature ($k=0,1,-1$) of the spatial sections $M_3$.
The $3-$space $M_3$ is usually taken to be one of the simply-connected
manilfolds: Euclidean $\mathbb{R}^3$, spherical $\mathbb{S}^3$, or
hyperbolic $\mathbb{H}^3$.
However, it is known that the great majority of such constant curvature
$3-$spaces are multiply-connected quotient manifolds of the form
$\mathbb{R}^3/\Gamma$, $\mathbb{S}^3/\Gamma$, and $\mathbb{H}^3/\Gamma$,
where $\Gamma$ is a fixed-point free group of isometries of the
corresponding covering space.
On the other hand, given that the connectedness of the spatial sections
$M_3$ has not been determined by cosmological observations, and
since geometry does not fix the topology, our $3$--dimensional space
may be one of these possible multiply-connected quotient manifolds
(for reviews on this issue see, e.g., the Refs.~\refcite{CosmTopReviews}).

The immediate observational consequence of a detectable nontrivial
topology\cite{TopDetec} 
of $M_3$ is the existence of the circles-in-the-sky, i.e., pairs of
matching circles will be imprinted on the cosmic microwave background (CMB)
anisotropy sky maps.\cite{Cornish1998Calvao2005}
Hence, to observationally probe a putative nontrivial topology of $M_3$,
one should examine the CMB maps in order to extract the pairs of correlated
circles and determine the spatial topology.

In this regard, in a few recent works\cite{Previous}
in the context of $\Lambda$ cold dark matter ($\Lambda$CDM) model,
it has been shown that the knowledge of the spatial topology
through the circles-in-the-sky offers an effective way of setting
constraints on the density parameters associated with matter ($\Omega_m$)
and dark energy ($\Omega_{\Lambda}$). In other words, it has been shown
in Refs.~\refcite{Previous} 
that specific circles-in-the-sky detectable
topology of the spatial section of the Universe can be used to reduce
the degeneracies in the density parameters plane $\Omega_m-\,\Omega_{\Lambda}$,
which arise from statistical analyses with data from current
observations (see also the related Refs.~\refcite{Related}).

The question as to whether the detection of a non-trivial cosmic topology
can also be used to set constraints on the equation of state parameters
naturally arises here.
In this paper we address this question by examining to what extent
a non-trivial spatial topology can be used to place further constraints
on the dark energy (DE) equation of state parameter $w_x$. To this end,
we shall assume the Poincar\'{e} dodecahedral space as the circles-in-the-sky
observable spatial topology, and reanalyze  the current constraints on
a two-parameter DE equation of state (EOS) of Ref.~\refcite{Jassal2005},
which arise from the Type Ia supernovae (SNe Ia) data from the Legacy
sample\cite{Astier2006} along with the baryon acoustic oscillations (BAO)
peak in the large-scale correlation function of the Sloan Digital Sky
Survey (SDSS),\cite{Eisenstein2005} and CMB shift parameter.\cite{Bond1997}

\section{Poincar\'{e} Dodecahedral Space}
\label{sec:pds}
Combining the first-year CMB  with other astronomical data, the WMAP
team\cite{Spergel2003} reported the estimated value of the total density
$\Omega_{\mathrm{tot}}=1.02 \pm 0.02$ ($1\sigma$ level), which includes
a positively-curved universe as a possibility. This reported sign for the
curvature density ($\Omega_{k}= 1- \Omega_{\mathrm{tot}} \leq 0$)
has been reinforced by the combination of  WMAP three year CMB measurements
with other observational data sets,  cf. Table~12 of Ref.~\refcite{Spergel2007},
wherein six different values for the $\Omega_{\mathrm{tot}}$ ranging from a
very nearly flat ($\Omega_{\mathrm{tot}}= 1.005^{+0.0060}_{-0.0068}$) to
positively curved ($\Omega_{\mathrm{tot}}= 1.023\pm 0.014$)
depending on the combination of data sets used to resolve the geometrical
degeneracy.

\begin{figure}[ht]
\centerline{\psfig{file=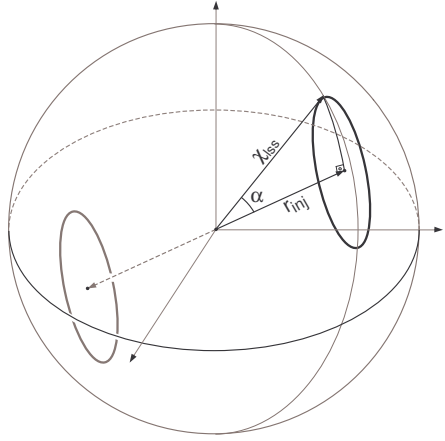,width=6.2cm}}
\vspace*{8pt}
\caption{A schematic illustration of a pair of antipodal matching
circles on the sphere of last scattering. The relation between
the angular radius $\alpha$ of the circles  and the angular
sides $r_\text{inj}$ and  $\chi^{}_\text{lss}$ is given by
Eq.(\ref{eq:cosalpha}).
\label{fig:circles}}
\end{figure}

In this work we assume the positively-curved Poincar\'e dodecahedral space (PDS)
as the circles-in-the-sky detectable topology of the spatial sections
of the Universe.\footnote{This topology is consistent with the observed total density,
and accounts for both the suppression of power of the low multipoles  of the first
and three year WMAP data\cite{Spergel2003,Spergel2007} and the temperature
two-point correlation function (see Refs.~\refcite{Luminet2003}--\refcite{Aurich2005a}).}
The fundamental polyhedron (FP) of the PDS topology is a regular spherical
dodecahedron that tiles the covering space $\mathbb{S}^3$ into 120
identical cells. The radius $r_\text{inj}$ (called the injectivity radius)
of the smallest sphere `inscribable' in the FP  is $\pi/10$.
The PDS is globally homogeneous, and a detectable
PDS topology ($\chi^{}_\text{lss} > r_\text{inj}\,$, cf.
Ref.~\refcite{TopDetec}) 
gives rise to six pairs of antipodal
circles-in-the-sky, one of which is shown in Fig~\ref{fig:circles}.

Now, a straightforward use of a Napier's rule to the right-angled
spherical triangle furnishes a relation between the angular radius
$\alpha$ and the angular sides $r_\text{inj}$ and
$\chi^{}_\text{lss}$ of the last scattering sphere, namely
\begin{equation}
\label{eq:cosalpha}
\cos \alpha = \frac{\tan r_\text{inj}}{\tan \chi^{}_\text{lss} }\;,
\end{equation}
which can be solved for $\chi^{}_\text{lss}$ to give
\begin{equation}
\label{eq:chilss}
\chi^{}_\text{lss} = \tan^{-1} \left[\,\frac{\tan r_\text{inj}}{\cos \alpha}\, \right].
\end{equation}
This equation shows that for a given topology, the measurement of a value
$\alpha$ for a circle-in-the-sky radius gives the distance $\chi^{}_\text{lss}$
to the last scattering surface in {\it units of the curvature radius} today,
$a_0=a(t_0)=(\,H_0\sqrt{|1-\Omega_{\mathrm{tot}}|}\,)^{-1}\,$.
Given a detection of a circle-in-the-sky, one can further constrain
the EOS parameters, as we shall show in Sec.~\ref{sec:ObsCons},
by comparing $\chi^{}_\text{lss}$ given by Eq.~\eqref{eq:chilss} with
that predicted by a model with specific EOS parametrization. 

\section{Dark Energy Parametrization}
\label{sec:de-param}

In the light of current observations, we assume the Friedmann-Lema\^{\i}tre%
-Robertson-Walker (FLRW) framework in which the current matter content of the
Universe is well approximated by a dust of density $\rho_m$ (baryonic plus
dark matter) along with a dark energy perfect fluid component of density
$\rho_x$ and pressure $p_x$. The Friedmann equation is then given by
\begin{equation}
\label{eq:friedmann}
H^2 =\frac{8 \pi G}{3c^2} (\rho_m + \rho_x) -\frac{kc^2}{a^2},
\end{equation}
where $H=\dot{a}/a$ is the Hubble parameter, overdot stands for
derivative with respect to time $t$, $G$ is Newton's constant, and
$c$ is the speed of light.

If one further assumes that these fluid components do not interact,
the energy conservation equations reduce to
\begin{equation} \label{ConsEq}
\dot{\rho}_i + 3\,H (\rho_i + p_i)=0,
\end{equation}
for each component $i=m, x$. This equation can
be rewritten in terms of the redshift $1+z = a_0/a$ and the
equation of state parameters $w_i = p_i/\rho_i$ in the form \begin{equation}
\frac{d\rho_i}{dz} = \frac{3\rho_i(1 + w_i)}{1+z},
\end{equation}
and has an integral solution
\begin{equation}
\rho_i (z) = \rho_i(0)\exp{\left[\int_0^z{\frac{3(1 + w_i)}{1+z'}\,\, dz'}\right]}.
\end{equation}

In this way, the Friedmann equation \eqref{eq:friedmann} can now be rewritten
in terms of the redshift $z$ and the dimensionless density variables
$\Omega_i=\rho_i(0)/\rho_{crit}(0)$, where $\rho_{crit}(z) = 3 c^2 H^2(z)/(8\pi G)$,
in the form
\begin{equation}
\label{eq:de-E2}
E^2(z) = \Omega_{m}(1+z)^{3} + \Omega_{k}(1+z)^{2} + \Omega_{x} f(z),
\end{equation}
where
\begin{equation}  \label{eq:f}
f(z) = \exp{\left[\int_0^z{\frac{3(1 + w_x(z'))}{1+z'}}\,\, dz'\right]},
\end{equation}
and where $E(z) = H(z)/H_0$ is the dimensionless Hubble function and
$\Omega_{k}=-kc^2 /(a_0^2 H_0^2)$ is the density curvature
parameter.

In this work we shall consider a model that is
defined by Eq.~(\ref{eq:de-E2}) and Eq.~(\ref{eq:f}) along with the
Jassal-Bagla-Padmanabhan\cite{Jassal2005} (JBP) dark energy
parametrization
\begin{equation} \label{JBP-param}
w_x(z) = w_0 + \frac{w_1 z}{(1+z)^2}\,,
\end{equation}
which combined with  Eqs.~(\ref{eq:de-E2}) and~(\ref{eq:f}) give
\begin{equation} \label{eq:Ez}
E^2(z) = \Omega_{m}(1+z)^{3}+\Omega_{k}(1+z)^{2}
+\Omega_{x}(1+z)^{3(1+w^{}_0)}\,\exp\left[\frac {3\, w_1 z^2}{2\,(1+z)^2}\right]\,.
\end{equation}
We note that for $z=0$ one has  the constraint equation
$\Omega_{x} = 1 - \Omega_{k} - \Omega_{m}$. 

{}From Eq.~(\ref{eq:chilss}) it is clear that to study the constraints
on the parameters of the JPB model we need the theoretical value
of $\chi^{}_\text{lss}$ predicted by this model, which is given
by
\begin{equation}
\label{eq:model_chi_lss}
\chi^{}_\text{lss} = \sqrt{|\Omega_k|} \,\,\,\frac{d_\text{c}(z^{}_\text{lss})}{d_H} =
\sqrt{|\Omega_k|}\int_0^{z_{lss}^{}} \frac{dz'}{E(z')},
\end{equation}
where $d_\text{c}(z) = d_H \int_0^z dz'/ E(z')$ is the comoving distance,
$d_H = c/H_0$ is the Hubble distance today, and $z^{}_\text{lss}$ is the
redshift of the last scattering surface.

\section{Observational Constraints}
\label{sec:ObsCons}
To illustrate the role played by the PDS topology in constraining the
EOS parameters of the model discussed in the previous section,
we have reanalyzed, with and without the topological contribution,
the constraints on the parameters that arise from Legacy
sample\cite{Astier2006} of SNe Ia along with  the baryon acoustic oscillations
(BAO) peak position in the galaxy power spectrum,\cite{Eisenstein2005} and
CMB shift parameter.\cite{Bond1997}

For the sake of brevity, we shall not discuss here the likelihoods
for the SNe Ia, BAO and CMB data, and simply  refer to them as
$\,L_\text{SN}(\vec{P})\,$, $\,L_\text{BAO}(\vec{P})\,$ and
$\,L_\text{CMB}(\vec{P})\,$ with $\vec{P} \equiv (\Omega_m, \Omega_k, w_0, w_1)$.
Further details on  these likelihoods can be found in a number
of papers as, for example, in Refs.~\refcite{Previous}b and
\refcite{Related}b. In this way, we focus on the likelihood of
topological origin, which is included in the statistical analysis
as a gaussian prior as follows.
The measurement  of a circle-in-the-sky of radius $\alpha$ in
CMB map involves unavoidably observational uncertainty $\sigma_{\alpha}$.
These observational pieces of topological information can be included
in the statistical analyses by using that $r_\text{inj} = \pi / 10$
(for PDS topology) along with Eq.~(\ref{eq:chilss}) to have $\chi^{}_\text{lss}$
and $\sigma_{\chi^{}_\text{lss}}$ (through the usual error propagation formula).
In this way, a detection of a PDS circles-in-the-sky would provide
an observational value of the radius of the last scattering surface and
the associated uncertainty, i.e. $\chi^{}_\text{lss} \pm \sigma_{\chi^{}_\text{lss}}$,
which we compare with the theoretical values predicted by the  JPB model by assuming a
gaussian distribution with the mean value given by Eqs.~\eqref{eq:Ez}
and~\eqref{eq:model_chi_lss}.
Thus, the likelihood associated to the topology is given by
\begin{equation}
L_\text{Top}(\vec{P}) \propto \exp{\left\{-\frac{1}{2} \,\,\frac{\,\,
\left[\chi_\text{lss}^{}
- \sqrt{|\Omega_k|} \,\,\, d_\text{c}^{}(z_\text{lss}^{}) \,\,d_H^{\,-\!1} \,\right]^2}
{\left(\sigma_{\chi_\text{lss}^{}}\right)^2} \right\}},
\end{equation}
whose dependence on the model parameters $\vec{P} = (\Omega_m, \Omega_k, w_0, w_1)$
clearly comes explicitly from $\Omega_k$ and from the comoving distance
$d_\text{c}(z^{}_\text{lss})$.  

\begin{figure*}[htb!]
\begin{center}
\includegraphics[scale=0.6]{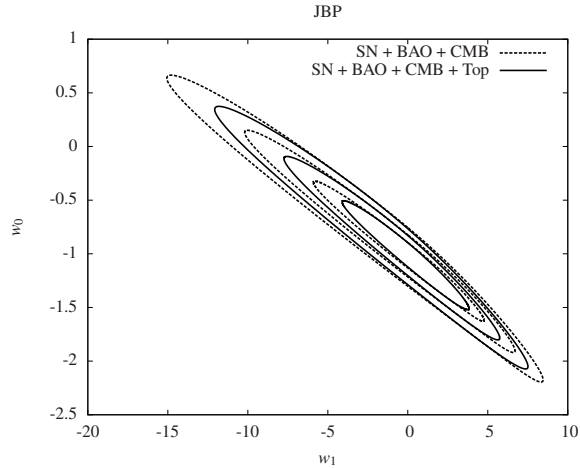}
\caption{Confidence regions for $1\sigma$--$\,3\sigma$ in the  $w_1 - w_0$ plane
for JBP parametrization. The dashed and full lines indicate
the confidence regions without  and with
the topological term $\,L_\text{Top}(\vec{P})\,$.
Clearly the PDS topology  reduces the degeneracies in the EOS parameters
$w_0$ and $w_1$ which arise from the SNe Ia, BAO and CMB.
\label{fig:w1_w0}}
\end{center} 
\end{figure*}

\section{Results and Conclusions}

In order to make apparent the role of the topological term $L_\text{Top}$ in
constraining the equation of state parameters, we have performed the joint
analysis maximizing the likelihoods with and without the topological term,
namely,
\begin{eqnarray}
\label{eq:likelihood}
L_1(\vec{P}) &=& L_\text{SN}(\vec{P})\,L_\text{BAO}(\vec{P})\,
L_\text{CMB}(\vec{P})\,L_\text{Top}(\vec{P})\,, \\
\nonumber \\
L_2(\vec{P}) &=& L_\text{SN}(\vec{P})\, L_\text{BAO}(\vec{P})\,
L_\text{CMB}(\vec{P}) \,,
\end{eqnarray}
where $L_\text{Top}(\vec{P})$ is given by a circle-in-the-sky
radius along with an unavoidable observational uncertainty in its
measurements, which we take to $\alpha = 20^{\circ} \pm 4^{\circ}$.
This value for the radius was not covered by the search for almost antipodal
correlated circles (with radii larger than $\alpha= 25^{\circ}$) made in Ref.~\refcite{Cornish-et-al-2004}, in which no circles were found.
\footnote{We note that another statistical technique for probing below
the limit of $20^{\circ}$ for a directed search for a \emph{specific}
topology was introduced in Ref.~\refcite{Key-et-al-2007}. By using
this combined search procedure with the knowledge of four or more
circle pairs, for example, one might be able to detect circles
with radii as small as $5^{\circ}$.}
We also used a conservative error of $20\%$ to account for the possible
uncertainties in the measurement of the circle radius.

Figure~\ref{fig:w1_w0} shows $1\sigma - 3\sigma$ confidence regions of
the JBP parameter $w_1 - w_0$ plane calculated by using the above
likelihoods $L_1(\vec{P})$ (with topology) and $L_2(\vec{P})$ (with no
topology) and marginalizing over the remaining parameters $\Omega_m$
and $\Omega_k$. It is clear from this figure that a circles-in-the-sky
detection of the PDS topology reduces appreciably the parametric
regions allowed by Legacy SNe Ia sample combined with  the BAO
peak position in the galaxy power spectrum, and CMB shift parameter.
In Tab.~\ref{tab:3s_limits} we explicitly quantify these constraints by
collecting together the $3\sigma$ lower and upper bounds of the
parameters $w_0$ and $w_1$ which were calculated without and with
the topological statistical term. Table~\ref{tab:3s_limits} makes
apparent that the allowed intervals for $w_0$ and $w_1$ are reduced,
respectively, of $\sim 14.4\%$ and $\sim 16.7\%$ by the topological term.

\begin{table}[ht!]
\tbl{The lower and upper $3\sigma$ limits of in the
$w_0-w_1$ plane without and with $L_\text{Top}$.}
{\begin{tabular}{@{}lcc@{}} \toprule
Model & $w_0$ & $w_1$
\\ \colrule
JBP & $(-2.194,0.665)$ & $(-15.059,8.457)$ \\
JBP+PDS & $(-2.073,0.374)$ & $(-12.069,7.510)$ \\ \botrule
\end{tabular}
\label{tab:3s_limits}}
\end{table}

\begin{figure*}[htb!]
\begin{center}
\includegraphics[scale=0.47]{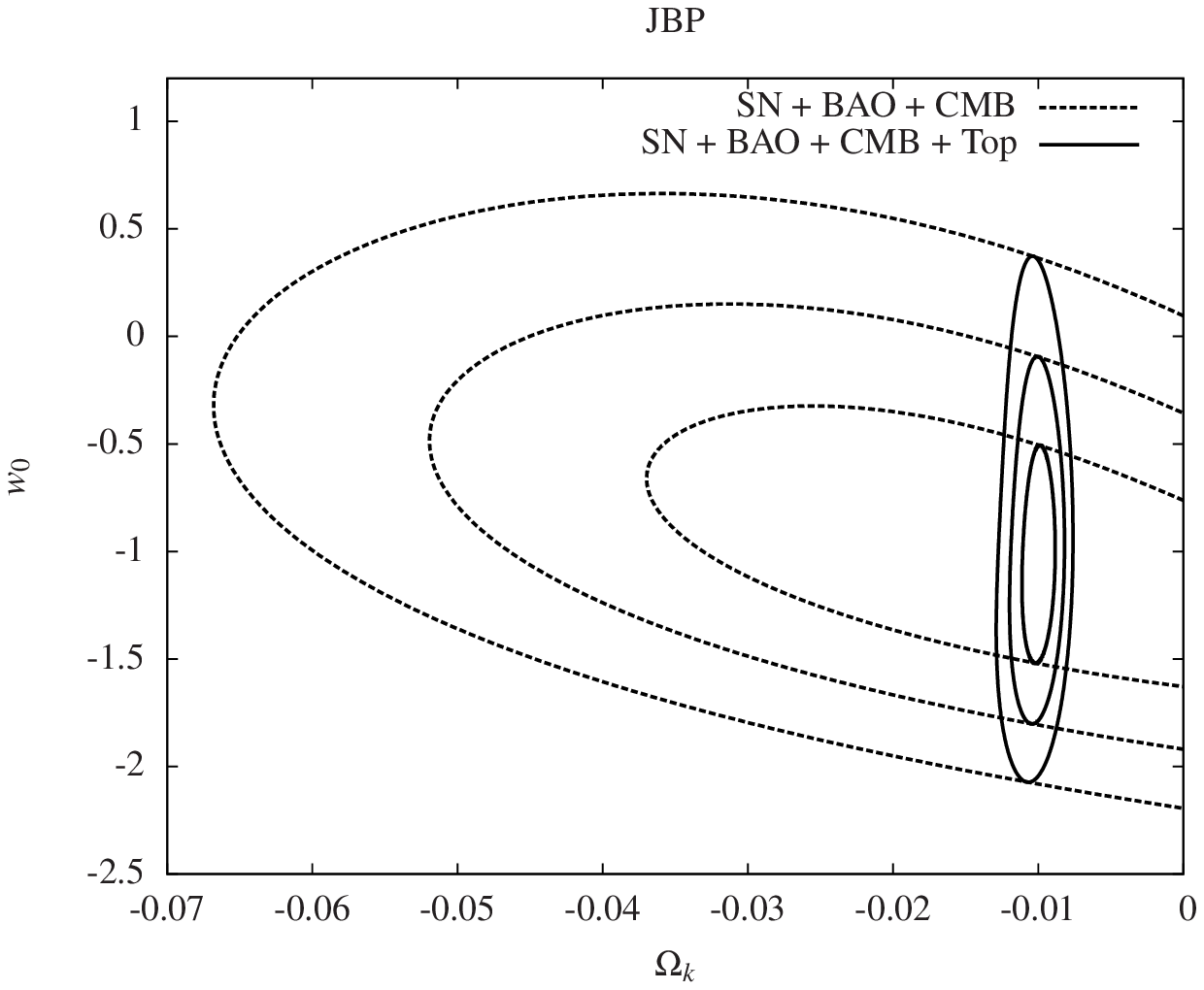}
\includegraphics[scale=0.47]{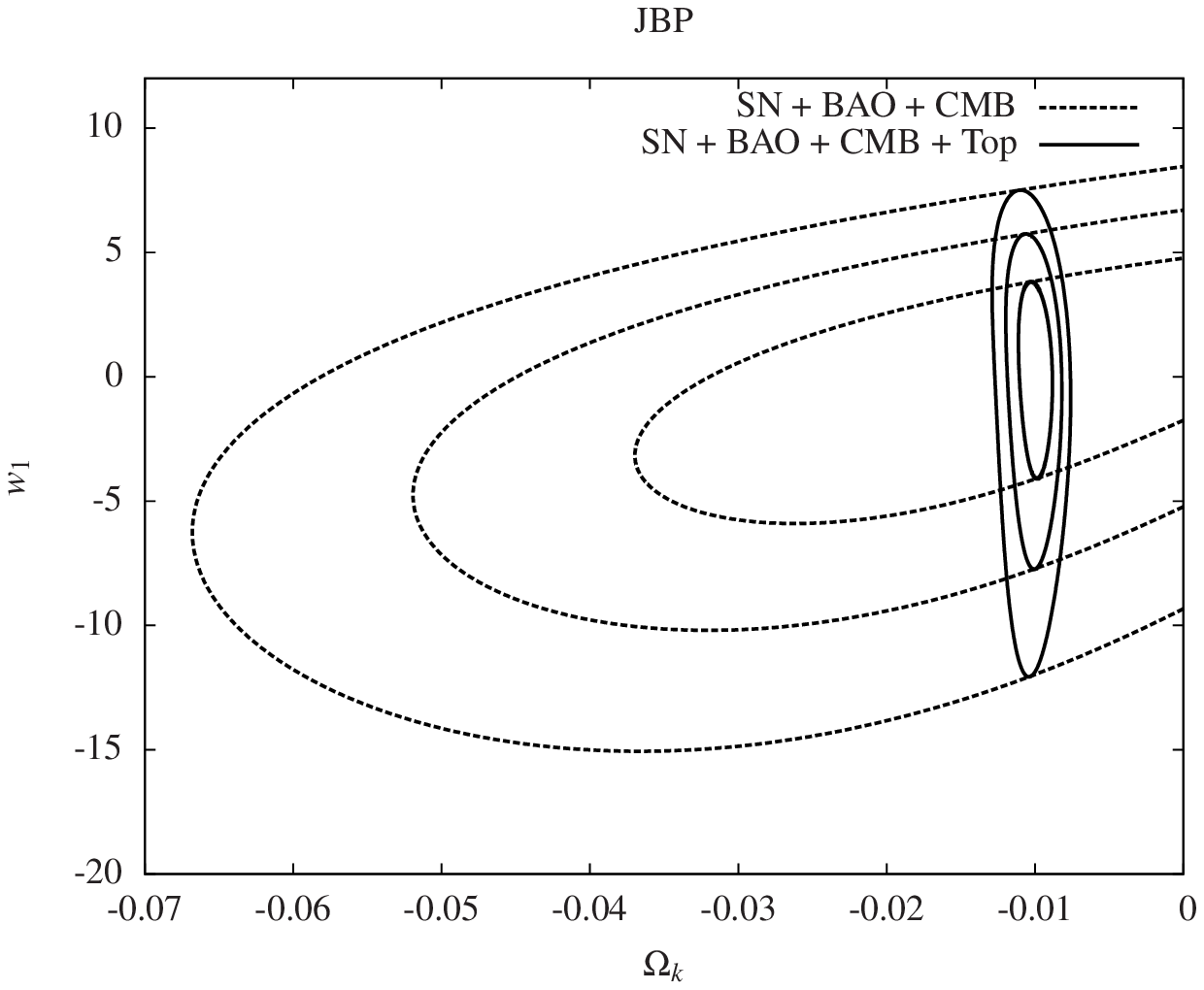}
\caption{Confidence regions in $1\sigma$--$\,3\sigma$ of $(\Omega_k, w_0)$
and  $(\Omega_k, w_1)$ for JBP  parametrization. The
main effect of the topological prior is to constrain the value
of $\Omega_k\,$.
\label{fig:Omega_k}  }
\end{center} 
\end{figure*}

To complete this study we show in Figure~\ref{fig:Omega_k} the
results of $1\sigma - 3\sigma$ statistical analysis for $\Omega_k - w_0$
and $\Omega_k - w_1$ planes (again marginalizing over the other
parameters). This figure makes apparent that the main role of the
topology in these parameter planes is to reduce considerably the
degeneracy on $\Omega_k$
allowed by the above-mentioned observational data, in agreement with
earlier related results obtained in different contexts.\cite{Related}
We also note that  the $w_0$ and $w_1$ $3\sigma$ limits are the same
on both Fig.~\ref{fig:w1_w0} and Fig.~\ref{fig:Omega_k}.
Finally, we note that it can be shown that a suitable Gaussian prior on the
curvature density parameter $\Omega_k$ can mimic the role of the topology
in such statistical analyses.

To conclude, we note that although we have illustrated the degree to which
a circle-in-the-sky topology detection reduces the degeneracies in the DE
EOS parameters  by using $\alpha = 20^\circ \pm 4^\circ$, our general
results should not change appreciably for smaller radius and
uncertainties, and its general features should hold
for other two-parameters EOS.

\section*{Acknowledgments}
M. J. R. acknowledges the support of FAPERJ under a CNE E-26/101.556/2010 grant.
This work was also supported by Conselho Nacional de Desenvolvimento
Cient\'{\i}fico e Tecnol\'{o}gico (CNPq) - Brasil, under grant No. 472436/2007-4.
M.J.R., S.V. and M.P.L. thank CNPq for the grants under which this work
was carried out.  We are also grateful to A.F.F. Teixeira for indicating
misprints and omissions.

\end{document}